# Digital Forensic Approaches for Amazon Alexa Ecosystem

Hyunji Chung[a], Jungheum Park[a], Sangjin Lee[a]

[a] Center for Information Security Technologies (CIST), Korea University, 145 Anam-ro, Seongbuk-Gu, Seoul, 08421, South Korea

**Abstract**

Internet of Things (IoT) devices such as the Amazon Echo - a smart speaker developed by Amazon - are undoubtedly great sources of potential digital evidence due to their ubiquitous use and their *always-on* mode of operation, constituting a human-life's black box. The Amazon Echo in particular plays a centric role for the cloud-based intelligent virtual assistant (IVA) Alexa developed by Amazon Lab126. The Alexa-enabled wireless smart speaker is the gateway for all voice commands submitted to Alexa. Moreover, the IVA interacts with a plethora of compatible IoT devices and third-party applications that leverage cloud resources. Understanding the complex cloud ecosystem that allows ubiquitous use of Alexa is paramount on supporting digital investigations when need raises. This paper discusses methods for digital forensics pertaining to the IVA Alexa's ecosystem. The primary contribution of this paper consists of a new efficient approach of combining cloud-native forensics with client-side forensics (forensics for companion devices), to support practical digital investigations. Based on a deep understanding of the targeted ecosystem, we propose a proof-of-concept tool, *CIFT*, that supports identification, acquisition and analysis of both native artifacts from the cloud and client-centric artifacts from local devices (mobile applications and web browsers).

## 1. Introduction

The Internet of Things (IoT) is evolving rapidly along with the network of physical objects that contain embedded communication technology. Analysts predict that the worldwide IoT market will grow to $1.7 trillion in 2020 with a compound annual growth rate (CAGR) of 16.9% [1]. Gartner predicts that 25% of households using an intelligent virtual assistant (IVA) will have two or more devices by 2020 [2]. The ubiquitous use of wearables, personal smart devices, smart appliances, etc., will generate a large amount of digital data that can be a great source of digital evidence.

In several recent criminal investigations, law enforcement officials, legal experts and forensics experts attempted to use "always-on" IoT devices as sources of forensic artifacts similar to human-life black boxes. In particular, one recent criminal investigation case involving an Amazon Echo, gained widespread attention in the media. In November 2015, James Bates was charged with first-degree murder of another man, who was found dead in Bates' hot tub. Police in Arkansas seized Bates' Alexa-enabled Echo smart speaker from his home, and asked Amazon to hand over any pertinent information regarding the device's communication with Alexa. However, Amazon denied the request in the absence of a valid and binding legal demand [3].

While there are many legal questions regarding the use of this type of evidence, there are also important technical considerations. Most importantly, to efficiently investigate these types of cases, it is first necessary to understand the digital forensic characteristics of Amazon's Alexa and its ecosystem.

When Alexa-enabled, Amazon Echo is not only a smart speaker, but operates as an intelligent, intelligent virtual assistant. As a cloud service, Alexa interacts with various Alexa-enabled devices such as Echo, and it can communicate with other compatible IoT devices and third-party applications by converting the voice requests to other services' native communication protocol. Also, for customizing these Alexa-related environments, users should access the cloud service using companion clients, such as PC or mobile (Android and iOS) devices. Thus, the ecosystem created by all these interconnected devices, third-party applications and companion clients is complex and heterogeneous [4]. In this paper, we will refer to this ecosystem as the *Amazon Alexa ecosystem*.

We propose a new digital forensic approach for the

---

Amazon Alexa ecosystem combining cloud-side and client-side forensics. The acquisition of cloud-native artifacts from the Alexa is very important. Echo operations are based on Alexa, so the cloud includes many different types of artifacts related to user behaviors. Unfortunately, this approach has two limitations. First, it requires valid user accounts in order to access the cloud. There is, of course, the potential to discover access information through investigation or interrogation, but this information is not always available. Second, it is difficult to recover deleted data on the cloud. Client-side forensics approaches are needed to overcome these limitations. That is, when it is impossible to acquire cloud-native artifacts, companion clients can offer important artifacts for digital investigations.

As a result of our analysis, we introduce a proof-of-concept tool for cloud-based IoT environments, *CIFT: Cloud-based IoT Forensic Toolkit*, which can acquire cloud native artifacts from Alexa using unofficial APIs and analyze client-side artifacts associated with the use of a web-based application. We also tried to normalize all identified artifacts into a database file, and visualize them for evaluating our approach and further supporting the work of the digital forensics community. In a situation where existing tools and procedures cannot meet the demand for this emerging IoT system, our findings and proof-of-concept tool will be helpful for investigators attempting to work in the Amazon Alexa environment.

The rest of the paper is organized as follows. Section 2 describes the target system and Section 3 reviews existing works. Section 4 presents our findings for digital forensics and Section 5 introduces an implementation based on our findings. Section 6 evaluates results with visualization techniques. Finally, Section 7 discusses conclusions and next steps.

## 2. Amazon Alexa and Digital Forensics

### 2.1. Research Motivation

In the IoT world, the do-it-yourself culture is encouraged, meaning users themselves can develop customized devices and applications for their IoT environments with tiny sensors and programmable brokers [5]. However, it is not easy for people who are unfamiliar with state-of-the-art technologies to build customized IoT environments. Thus, most people tend to purchase IoT consumer products, including but not limited to smart assistants, lights, sensors, switches, hubs, thermostats, and fitness devices.

Although a variety of products are available on the market, we focused on one of the most famous products, Amazon Echo. The Amazon Echo family of smart devices, which also includes Dot and Tap, connect to the intelligent cloud-based voice service, Alexa Voice Service (AVS). With Alexa as a voice-activated personal assistant, the Echo is capable of doing various things, such as managing to-do lists, playing music, setting alarms, placing orders, searching information, and controlling other smart devices [4]. According to an industry report, the Echo family sold more than 11 million units between the middle of 2015 and 2016 [6]. Additionally, as announced at CES 2017, there is an interesting convergence of the Alexa with various devices, such as connected cars, smartww fridges, and robots, which indicates that the Amazon Alexa-related environment will become an important source of potential digital evidence. For these reasons, the Echo and Alexa were selected as the first targets for studying digital forensic approaches inside the IoT world.

### 2.2. Amazon Alexa Ecosystem

Before presenting our analysis, we describe the detailed architectures related to the target IoT environment. As mentioned above, the Amazon Echo controls an interface for communicating with the cloud-based service, Alexa. Cloud-based operations, such as Echo and Alexa, represent a general operating method of IoT consumer products because most are inseparable from cloud services in providing interoperability with companion clients and compatible devices for user convenience. Therefore, this subsection describes the target IoT environment focusing on its cloud service, Alexa.

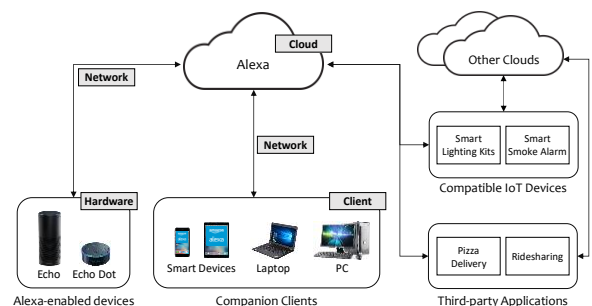
Figure 1. Amazon Alexa ecosystem

The Amazon Alexa ecosystem is composed of various components, as shown in **Fig. 1**. First, one or more Alexa-enabled devices are required for talking to the Alexa cloud service. We should note our description specifically relates to the ecosystem associated with Amazon's Echo devices, and other devices exist for communicating with the cloud service. Next, the Alexa in the figure represents all Amazon cloud platforms supporting operation of this ecosystem. So it includes various cloud services for authentication, data management, and logging, as well as the Alexa Voice Service. In addition to these core components, one of the interesting aspects from the viewpoint of digital forensics is that companion clients are essential to managing the overall operating environment through access to the cloud server. The companion client means a personal device for executing Alexa companion applications, such as the Amazon Alexa App for Fire OS, Android, and iOS. Although there is no official application for PC, users can utilize web browsers for accessing the cloud, so any kind of digital device with web browsing capability may have potential digital evidence relating to the use of the Alexa ecosystem. Furthermore, Alexa can be expanded through connections to compatible IoT devices and through addition of skills (third-party apps) for utilizing various services, as shown in **Fig. 1**.

*2.3. Analysis strategy*

Due to the characteristics of the Alexa ecosystem introduced above (summarized in **Table 1**), we considered multi-level forensic approaches similar to that proposed by Zawoad et al. [7]. This subsection outlines multi-level forensic approaches for the target ecosystem and also presents the scope of the study.

*2.3.1. Hardware: Alexa-enabled devices*

Each Alexa-enabled device needs to be decomposed for performing hardware-level analysis. In this light, Clinton et al. introduced their research on analyzing Amazon Echo at the hardware level [8]. The authors summarized their experiments for reverse engineering the device through available approaches, such as eMMC Root, JTAG, and debug ports. Although they explained some possible methods for enabling access to the internal parts, including soldered memory chips, the authors did not mention details about data stored within the device. So, this viewpoint will be included in one of our future projects, expanding on the work presented here.

*2.3.2. Network: Communication protocol*

As marked in **Fig. 1**, Alexa-enabled devices and companion clients should communicate with the Alexa through the internet. As a result of traffic analysis using the Charles web debugging proxy [9], we confirmed that most traffic associated with forensically meaningful artifacts are transferred over an encrypted connection after creating a session with a valid user ID and password. With this network analysis, we were able to efficiently identify cloud-native and client-centric artifacts. The results will be described in Section 4 in detail.

*2.3.3. Cloud: Alexa cloud service*

As mentioned in Section 2.2, the Alexa is a core component of the target ecosystem. Like other cloud services, Alexa operates using pre-defined APIs to transceive data, but unfortunately the available API list is not officially open to the public. As far as we know, there is no literature for acquiring native artifacts from Alexa from the viewpoint of digital forensics. Thus, we performed an intensive analysis to reveal unofficial APIs used by Alexa and to acquire cloud-native artifacts for supporting investigations.

*2.3.4. Client: Alexa companion clients*

Lastly, there is one more level of analysis relating to Alexa companion clients. Interestingly, the use of at least one companion client is essential to set up Alexa-enabled devices and manage the operating environment. For example, users can configure environment settings, review previous conversations with Alexa, and enable/disable skills using a mobile app or web-browser. In this process, a large amount of data associated with accessing Alexa can be stored naturally in companion clients. This makes it necessary to acquire these client-centric artifacts and consolidate them along with cloud-native artifacts. For support of that effort, we attempted to discover all available artifacts from both the cloud and client level.

Table 1. Analysis approaches for Amazon Alexa ecosystem

| Level | Description | Progress |
|---|---|---|
| Cloud | - Acquiring cloud native artifacts from Alexa (when user credentials are available) | Section 4 describes the results |
| Client | - Identifying client-centric artifacts from mobile apps and web-browsers | Section 4 describes the results |
| Network | - Understanding the communication protocol and unofficial APIs used by Alexa | Section 4 describes the results |
| Hardware | - Reveling data stored in Alexa-enabled devices | Future work |

## 3. Related works

*3.1. IoT and digital forensics*

There have been previous studies of digital forensic frameworks for the IoT environment. Oriwoh et al. introduced hypothetical IoT crime scenarios that were carried out by a suspect who used various IoTware to commit crimes. Based on these scenarios, the authors also discussed potential sources of digital evidence [10]. Hegarty et al. and Liu established some fundamental overarching challenges in the four main phases (identification, preservation, analysis, and presentation) of digital forensics and identified key issues for IoT forensics [11, 12]. Zawoad et al. defined IoT forensics as a combination of device, network, and cloud forensics [7]. Also, Kebande et al. proposed a framework for digital investigations in the IoT domain and compared it with existing models. Their framework provided a combination of three distinct processes, which included the proactive process, IoT forensics, and the reactive process [13].

Previous studies have proposed generic frameworks for defining theoretical models for the IoT environment. With these studies as starting points for understanding the IoT world, we present a practical digital forensic approach for the Amazon Alexa ecosystem that goes one step further.

*3.2. Cloud forensics*

As mentioned in Section 2, cloud forensics should play a key role in approaching the Amazon Alexa ecosystem as a source of digital evidence. Existing studies have proposed two perspectives: client-based cloud forensics and cloud-native forensics.

In the first stage of research on cloud forensics, many researchers have performed client-based cloud forensics, acquiring and analyzing data that was locally saved by applications or web browsers related to the use of famous cloud services, including but not limited to Amazon S3, Google Docs, Dropbox, Evernote, and ownCloud [14, 15, 16, 17, 18].

Next, recent research efforts have proposed examining cloud-native forensics to overcome the fundamental limitation that there is still large amounts data not stored in storage devices or just stored in temporary caches. For example, Vassil et al. introduced forensic approaches on native artifacts of Google Drive, Microsoft OneDrive, Dropbox, and Box using APIs supported by the cloud services. As a result, the authors were able to implement cloud drive acquisition tools [19, 20].

*3.3. Preliminary studies on Amazon Alexa forensics*

There have been several efforts to identify forensic artifacts on IoT consumer products. For instance, a research team briefly presented their findings on artifacts saved when users utilize Android mobile apps relating to IoT products, such as WinkHub, Amazon Echo, Samsung SmartCam, and Nest products [21]. In particular, regarding Amazon Echo, the team mentioned SQLite databases and web cache files that include meaningful information, such as accounts and interactions with Alexa. Although their findings include interesting data acquired by client-based cloud forensics, the work was only partially useful for digital forensics of the entire ecosystem. Similarly, Benson posted a python script to parse a SQLite database file of an Amazon Alexa app for iOS backed up via iTunes [22]. The python script was able to parse out objects in to-do and shopping lists.

*3.4. Research direction affected by related works*

As a result of our literature review, we decided to combine two perspectives on cloud forensics in order to propose an integrated IoT forensic system for the Amazon Alexa ecosystem. In detail, cloud-native forensics is essential for identifying user behaviors, because most meaningful data are only saved on the cloud side. However, acquiring data from the cloud has two fundamental limitations in that it requires valid user credentials (usually a set of ID and password), and it is practically impossible to respond to a situation in which users try to delete data from the cloud. To handle these limitations in practice, it is necessary to find client-centric artifacts stored within companion clients, and, if available, use them to enhance results of cloud-native forensics. Furthermore, because the client-centric artifacts, including local databases and cache files, are also closely related to the cloud-side data, it is necessary to first understand the raw cloud data.

**4. Forensic Artifacts on Amazon Alexa Ecosystem**

This section describes forensic artifacts from the use of Alexa-enabled devices and companion clients.

*4.1. Test environment*

To identify meaningful artifacts on the target ecosystem, a test environment was established, as listed in **Table 2**. In this environment, we performed experiments for about two months with two Amazon Echo Dot products. This work also tested a variety of companion clients associated with Android, iOS, OS X, and Windows. In detail, the first two mobile devices were used for testing the Amazon Alexa app, and the other clients were utilized to access the Alexa web server using popular web browsers.

Table 2. Test environments

| Item | Description |
|---|---|
| Alexa-enabled devices | (1) Echo Dot (S/N: ***0L9***473***P) <br> (2) Echo Dot (S/N: **90***964*****U) <br> * some characters of S/N are masked by asterisks |
| Companion clients and applications | (1) Android 4.4.2 + Alexa app (1.24.1176.0) <br> (2) iOS 10.1.1 + Alexa app (1.24.1176.0) <br> (3) OS X 10.10.5 + Chrome (55.0.2883.87) <br> (4) Windows 10 + Chrome (55.0.2883.87) |
| Total test period | 2016-11-18 ~ 2017-01-29 |
| Last verification date | 2017-01-29 |

All findings introduced in this section were verified through repetitive tests conducted until Jan. 29, 2017. We should note that although the tests included popular web browsers, such as Internet Explorer, Edge, and Safari, they were excluded from the results of this paper because they did not store any meaningful cache data in the use of the Alexa web application.

*4.2. Cloud native artifacts*

The key to an understanding the target ecosystem is identifying native data stored on the cloud side.

*4.2.1. Revealing unofficial Alexa APIs*

The first attempt to go inside the target was to understand the communication protocols and data formats used by Alexa. For support of this work, we performed an intensive traffic analysis using a web proxy. Based on the analysis, we confirmed that most traffic is transferred over encrypted connections, and native artifacts are returned as JSON format.

The next step was to reveal how to get specific data from the cloud on the assumption that we have valid user credentials. Like other cloud services, the Alexa cloud service also operates with pre-defined APIs to transceive data. Although the API list is officially not open to the public, there have been efforts to reveal unofficial APIs of Alexa [23, 24]. In addition to this information, we identified additional useful APIs and interpreted their return values from the cloud in order to discover configurations and user activities.

*4.2.2. Understanding Alexa's native artifacts*

**Appendix A** lists our findings. The second column of the appendix shows unofficial Alexa APIs identified and used in this study. For ease of terminology, each API has a name highlighted in bold. To clarify our findings, **Fig. 2** shows the relationship between menus of the web-based application for Alexa and APIs listed in the appendix. Also, regarding actual JSON data returned by APIs, the last column of the table summarizes important keys and values that describe the meaning of APIs in summary.

As a result of data analysis, each API is categorized as one of the following seven categories: account, customer setting, Alexa-enabled device, compatible device, skill, user activity, and etc. These categories are closely connected with the data normalization strategy (Section 5). As shown in **Appendix A**, we can acquire forensically meaningful native artifacts from the Alexa, such as registered user accounts, Alexa-enabled devices, saved Wi-Fi settings (including unencrypted passwords), linked Google calendars, and installed skill lists that may be used to interact with other cloud services.

One of the particularly interesting things we found is that there is a large amount of data with timestamps. More specifically, JSON data acquired by APIs such as *cards*, *activities*, *media*, *notifications*, *phoenix*, and *todos*, contained values with UNIX timestamps. This may provide sources of evidence that allow reconstruction of user activities with a time zone identified by *device-preference* API. In addition, there were also interesting values acquired by *cards*, *activities*, and *todos* that included the rear part of a URL possessing a user's voice file on the cloud. Thus, it is possible to download the voice file using *utterance* API if necessary.

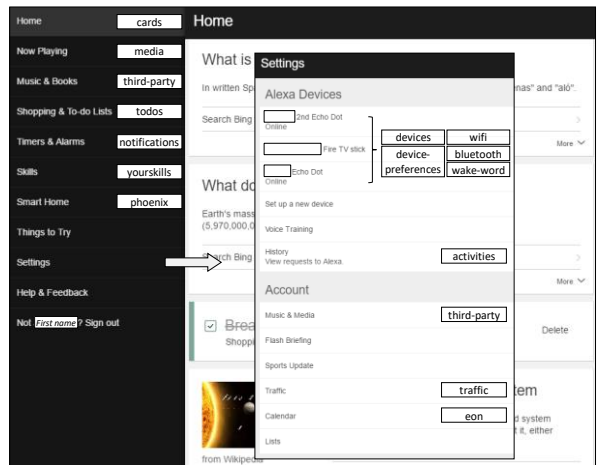

Figure 2. Web-based application for Alexa

*4.3. Client-centric artifacts*

After understanding Alexa's native artifacts, we tried to find out client-centric artifacts from the use of the companion applications listed in **Table 2**. To sum up our findings, **Table 3** provides locations of artifacts having Alexa data in the companion devices.

*4.3.1. Databases of the Alexa mobile app*

With smart devices based on Android and iOS, users can manage their own Alexa ecosystem through an official mobile application developed by Amazon like shown in **Fig. 2**.

In the Android system, the application uses two SQLite files: map_data_storage.db and DataStore.db. The first database contains token information about a user who is currently logged in. That is, all data in this database are deleted when a user signs out. Of course, a part of the deleted records could be found from unused areas of the SQLite database and its journal file, but this study did not cover a situation where data was deleted. In addition, the other file includes to-do and shopping lists. These lists can be acquired from the cloud using *todos* API, as described in **Appendix A**.

In the case of the iOS system, the application manages one SQLite file titled LocalData.sqlite. This file also includes to-do and shopping lists like DataStore.db in Android [22]. We should note that we solely analyzed files acquired by the iTunes backup protocol. So, there was a limitation on accessible files relating to the target application.

Table 3. Client centric artifacts from companion clients discussed in this paper

| OS | Application | Path | Format | Description |
|---|---|---|---|---|
| Android 4.4.2 | Alexa 1.24.1176.0 | /data/data/com.amazon.dee.app/databases/map_data_storage.db | SQLite | Tokens of an active user |
| | | /data/data/com.amazon.dee.app/databases/DataStore.db | SQLite | Todo and shopping list |
| | | /data/data/com.amazon.dee.app/app_webview/Cache/* | WebView cache | Cached native artifacts |
| iOS 10.1.1 | Alexa 1.24.1176.0 | [iTunes backup]/com.amazon.echo/Documents/LocalData.sqlite | SQLite | Todo and shopping list |
| OS X 10.10.5 | Chrome 55.0.2883.87 | ~/Library/Caches/Google/Chrome/Default/Cache/ | Chrome cache | Cached native artifacts |
| Windows 10 | Chrome 55.0.2883.87 | %UserProfile%\AppData\Local\Google\Chrome\User Data\Default\Cache\ | Chrome cache | Cached native artifacts |

The results of database file examinations showed us that there was little information stored locally on companion devices by the application. Meanwhile, although XML or PLIST files also exist for storing preferences, they were not in used in this study.

*4.3.2. Android WebView cache*

As the Amazon Alexa is basically a web-based application, it uses the WebView class to display online content in Android [25]. Thus, there is a chance that cloud-native artifacts are cached by the WebView.

The cache directory described in **Table 3** may contain multiple cache files. Each WebView cache file (of Android 4.4.2) simply consists of a string with the original URL and a data stream. Based on existing comments [21, 26], **Fig. 3** (a) describes the internal format of the WebView cache. We found an 8-byte fixed header and footer, and also a 4-byte field for storing the length of a string with the original URL. As an example of Alexa-related caches, **Fig. 3** (b) shows a file cached after calling *phoenix* API. Because a data stream of the example is gzip-compressed data, it was necessary to decompress it to obtain the original JSON.

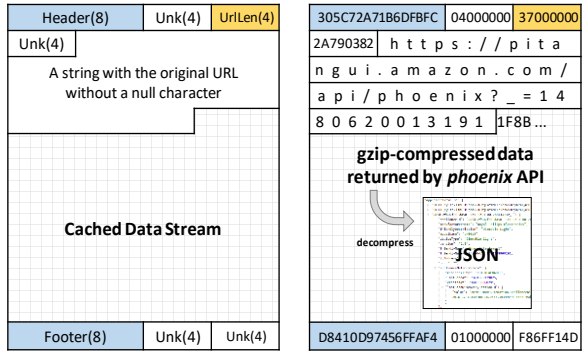

(a) WebView Cache Internals  (b) Alexa cache as an example
Figure 3. Android WebView cache format

*4.3.3. Chrome web cache*

As an alternative to the mobile app, users can utilize any web browser for accessing the Alexa website, https://alexa.amazon.com (= pitangui.amazon.com).

After performing experiments with popular web browsers, we confirmed that there is a possibility of acquiring cloud-native artifacts from local companion clients cached by Google Chrome. Because the format of Chrome's disk cache is already well known, this paper outlines practical approaches to acquiring Alexa-related data from the cache structure.

In the Chrome cache system, the data stream with Alexa native artifacts is usually stored inside the data block files (data_#) for small amounts of data. Thus, it was necessary to parse all cache entries and verify their data streams inside the data block files. Although the data stream may be stored as a separate file, there is a small possibility because it is a compressed JSON string with gzip. Like normal cache entries, Alexa-related caches have two data streams: the first one is for HTTP headers, and the other is actual cached data.

These cache data, both Android WebView and Chrome web caches, have great potential as sources of digital evidence. Although these caches are very helpful for understanding user behaviors, especially when valid user credentials are not available or some native artifacts are deleted from the cloud, it also has inevitable limitations. For example, caches are only created when users click menus that trigger Alexa APIs, and they also can be deleted or overwritten at any time.

## 5. Design and Implementation

This section explains the overall design concept and implementation of an integrated system to support digital investigations on the Amazon Alexa ecosystem.

*5.1. Design of CIFT*

The *CIFT* (Cloud-based IoT Forensic Toolkit) is designed based on forensic approaches proposed in this paper. The *CIFT* provides a common interface for users to execute forensic components implemented for various IoT consumer products. As an example, this paper introduces a component for the Alexa ecosystem. **Fig. 4** shows the event flow diagram of an *CIFT* component for Alexa. As shown in the figure, the component is composed of four modules: UIM, CNM, COM, and DPM.

The user interface module (UIM) provides basic interface methods for setting up environments, as well as adding and processing user inputs. Each user input should consist of an operation type and pre-defined argument list for the operation. If an input is assigned with an operation type relating to the cloud native, the

UIM requests processing the input to the cloud-native module (CNM). In contrast, if an input is related to companion clients, the companion client module (CCM) involves the following process.

To run the CNM, it is first necessary to create a web session with the target cloud system (Alexa in this paper) using an ID and password passed as arguments. After a successful login, the CNM tries to acquire (download) cloud-native artifacts from the server using the unofficial APIs introduced in Section 4. When the server returns JSON data, the CNM requests parsing them to the data-parsing module (DPM). The DPM is in charge of parsing Alexa-related data, storing original files to the evidence library, and inserting normalized records into a database file.

In the case of processing user inputs for companion clients, the CCM operates internal methods according to a selected operation type, including but not limited to the Alexa app for Android, Alexa app for iOS, and the Chrome cache. For example, if the operation type is Alexa app for Android, a remaining argument should have a path to the root of the application directory. Thus, the CCM can acquire client-centric artifacts from the databases or cache files introduced in Section 4. Afterward, the following procedures are similar to flows of the CNM that include parsing raw data and saving normalized records using the DPM.

### 5.2. Implementation

#### 5.2.1. CIFT for Alexa

We developed a python package based on the proposed design concept. **Table 4** shows a sample program using Alexa modules of *CIFT*. As shown in the table, the UIM provides three primary methods for setting up environments (result directory and browser driver) and adding/processing user inputs.

For the verification and improvement, the source code will be released to the public in the near future.

Table 4. An example of executing Alexa modules of *CIFT*

| Step | Sample codes |
|---|---|
| Create an instance | `amazon_alexa = AmazonAlexaInterface()` |
| Set up environments | `amazon_alexa.basic_config(`<br>    `path_base_dir="~/CIFT-Result/",`<br>    `browser_driver=CIFTBrowserDrive.PHANTOMJS)` |
| Add inputs | `amazon_alexa.add_input(`<br>    `CIFTOperation.CLOUD,`<br>    `"**n*h**@*mail.com", "my@1234#password")`<br>`amazon_alexa.add_input(`<br>    `CIFTOperation.COMPANION_APP_ANDROID,`<br>    `"~/Alexa/Android/com.amazon.dee.app/")`<br>`amazon_alexa.add_input(`<br>    `CIFTOperation.COMPANION_APP_IOS,`<br>    `"~/Alexa/iOS/com.amazon.echo/")`<br>`amazon_alexa.add_input(`<br>    `CIFTOperation.COMPANION_BROWSER_CHROME,`<br>    `"~/Alexa/Windows 1/chrome_cache/")`<br>`amazon_alexa.add_input(`<br>    `CIFTOperation.COMPANION_BROWSER_CHROME,`<br>    `"~/Alexa/Windows 2/chrome_cache/")` |
| Process inputs | `amazon_alexa.start()` |

#### 5.2.2. Data normalization

As mentioned above, the results of *CIFT* are saved in a database file. To support efficient analysis, we proposed a normalized database schema for managing Alexa-related artifacts from both cloud and local systems. The current implementation supports our normalization strategy, as shown in **Appendix A** and **B**. In particular, we tried to normalize forensic artifacts with timestamps based on the l2t_csv (by log2timeline) format. **Appendix B** shows an example of records stored in the TIMELINE table**.**

A sample database created by *CIFT* will be explained using some visualization techniques in the next section.

### 6. Visualization and Evaluation

This section demonstrates the usefulness of *CIFT* through visualization. Although there are of course various ways to achieve this purpose, as an example, this paper utilizes a famous data processing environment called Elastic Stack (also known as ELK Stack), which is a group of open source products including Elasticsearch, Logstash and Kibana in order to search, analyze and visualize data in real time [27].

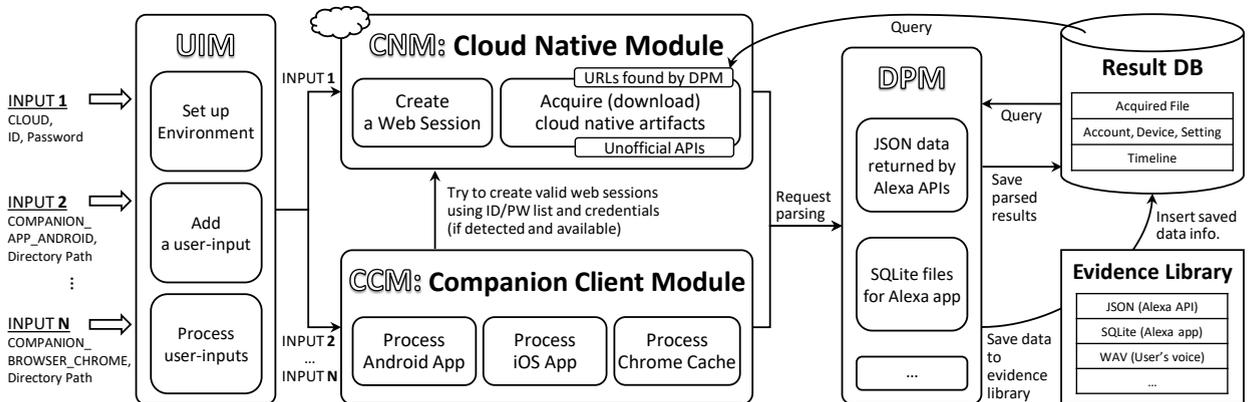

Figure 4. Event flows between Alexa modules of *CIFT*

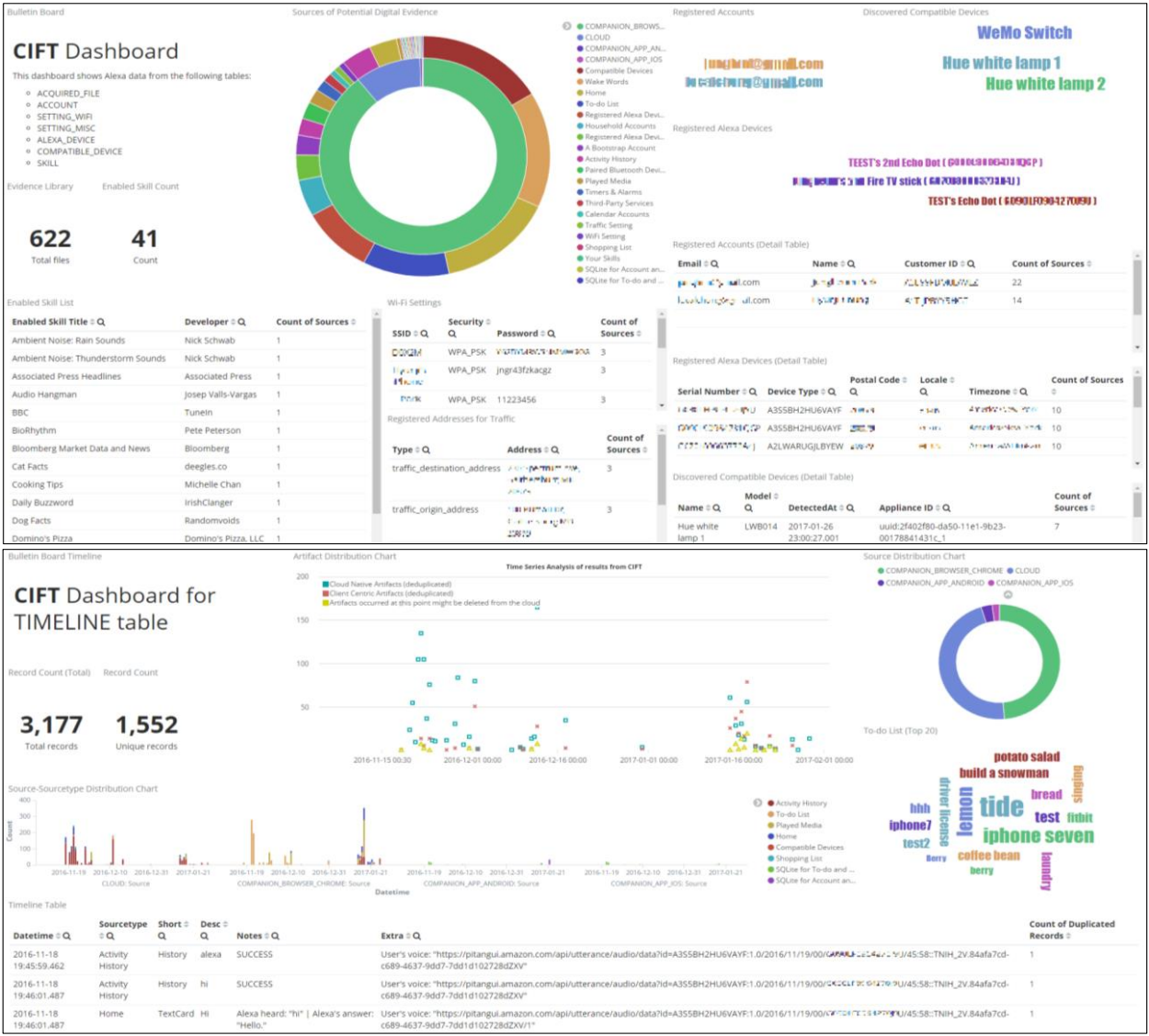

Figure 5. Visualization results of a database created by *CIFT*

The **Fig. 5** shows two dashboards designed for quick and efficient data analysis. The first dashboard is created for displaying data stored within all tables except for TIMELINE table. On the left side of this dashboard, two numbers show total files saved in an evidence library and skills enabled in Alexa respectively. More specifically, a pie chart displays detailed sources of potential digital evidence saved in the evidence library. As a multi-series donut chart, the inner circle represents operation types of *CIFT* as used in **Table 4**, and the outer circle shows types of data sources identified by each operation. In addition, this dashboard provides various data tables that list meaningful information such as enabled skills, Wi-Fi settings, user accounts, and compatible devices.

The second dashboard is designed exclusively for TIMELINE table. To provide an overview of artifacts, the top-middle chart displays various distributions that show when events relating to cloud native (square) and client centric (cross) artifacts occurred. The triangle symbols in this chart mean that artifacts occurred at these points might be deleted from the cloud, but they are identified from companion clients. There are also two additional charts to support data analysis by subdividing artifacts into groups that represent sources of potential digital evidence. In addition to these charts, this dashboard includes a data table for actual records, and a tag cloud showing to-do and shopping items.

Although it is hard to objectively evaluate forensic approaches proposed here due to a lack of related studies, these visualization results support the claim that they will be useful and helpful for investigations.

For future improvements, a Logstash config file for processing data and two Kibana dashboards (including visualization objects) will be provided together with the source of *CIFT*.

# 7. Conclusion and Future works

Currently, IoT devices are evolving rapidly and spreading widely in our lives. Many people are becoming accustomed to interacting with various IoT consumer products, such as virtual assistants, smart home kits, and fitness trackers. In these circumstances, lots of data are being produced in real time in response to user behaviors. Interestingly, these kinds of products are closely connected with both cloud and local systems, so it is possible to conduct integrated analysis of forensically meaningful data from both systems upon consideration of the target device's ecosystem.

Until now, there has been little research reported on Amazon Echo and its ecosystem in terms of digital forensics. As part of an effort to prepare for IoT, we proposed a new integrative approach combining cloud-native and client-centric forensics for the Amazon Alexa ecosystem. In addition, we introduced an implementation, *CIFT*, to acquire native artifacts from Alexa and analyze local artifacts from companion clients. Our findings and a tool developed based on the research will be valuable for digital investigations.

We have future plans for enhancing the results of this study, including but not limited to approaching the hardware level of Alex-enabled devices and performing memory forensics for delving into volatile artifacts. In addition, we will also expand our research with various IoT consumer products and implement new components of *CIFT* in order to support forensic activities relating to cloud-based IoT environments.

Appendix A. Unofficial Alexa APIs

| Category | Unofficial Alexa APIs ('{}' should be filled with appropriate values) | Description | Data Normalization (Refer to Appendix B) | Brief excerpts from JSON data returned by APIs (Personal info. is masked by asterisks and italic text) |
|---|---|---|---|---|
| Account | https://pitangui.amazon.com/api/**bootstrap** (https://pitangui.amazon.com/api/authentication) | Primary customer info. | ACCOUNT | "authentication": {<br>  "customerEmail": "*email address*",<br>  "customerId": "*customer id*",<br>  "customerName": "*customer name*"<br>} |
| Account | https://pitangui.amazon.com/api/**household** | Household accounts | ACCOUNT | {<br>  "email": "*email address*",<br>  "fullName": "*full name*",<br>  "id": "\*\*E99\*\*\*0L\*\*\*Z"<br>} |
| Alexa-enabled Device | https://pitangui.amazon.com/api/**devices**/device | Alexa devices such as Echo, Dot and Fire TV | ALEXA_DEVICE | {<br>  "deviceOwnerCustomerId": "\*\*E99\*\*\*0L\*\*\*Z",<br>  "deviceType": "A3S5BH2HU6VAYF",<br>  "macAddress": "\*\*71\*\*82\*\*FD",<br>  "serialNumber": "\*\*90\*\*\*964\*\*\*\*\*U",<br>  "softwareVersion": "564196920"<br>} |
| Alexa-enabled Device | https://pitangui.amazon.com/api/**device-preferences** | Preferences on registered Alexa devices | ALEXA_DEVICE | {<br>  "deviceAddress": "*registered address*",<br>  "deviceSerialNumber": "\*\*90\*\*\*964\*\*\*\*\*U",<br>  "locale": "*locale*", "postalCode": "*postal code*",<br>  "timeZoneId": "*time zone*"<br>} |
| Customer Setting | https://pitangui.amazon.com/api/**wifi**/configs | Wi-Fi settings | SETTING_WIFI | "values": {<br>  "ssid": "*SSID*", "preSharedKey": "*password (plain text)*",<br>  "securityMethod": "WPA_PSK"<br>} |
| Customer Setting | https://pitangui.amazon.com/api/**bluetooth** | Paired Bluetooth devices | SETTING_MISC | {<br>  "address": "*mac address*",<br>  "friendlyName": "*device name*"<br>} |
| Customer Setting | https://pitangui.amazon.com/api/**traffic**/settings | Location information for the traffic update | SETTING_MISC | "destination": { "label": "*address of a location*" },<br>"origin": { "label": "*address of a location*" } |
| Customer Setting | https://pitangui.amazon.com/api/**wake-word** | Wake word list for Alexa devices | SETTING_MISC | {<br>  "deviceSerialNumber": "\*\*90\*\*\*964\*\*\*\*\*U",<br>  "wakeWord": "ALEXA"<br>} |
| Customer Setting | https://pitangui.amazon.com/api/**third-party** | Third party services | SETTING_MISC | {<br>  "associationState": "ASSOCIATED",<br>  "serviceName": "*service name*"<br>} |
| Customer Setting | https://pitangui.amazon.com/api/**eon**/householdaccounts | Linked Google calendars | SETTING_MISC | {<br>  "emailId": "*google account*",<br>  "calendarList": [{<br>    "calendarId": "*calendar id*", "calendarName": "*name*"<br>  }]<br>} |
| Skill | https://skills-store.amazon.com/app/secure/**yourskills**<br>* A HTTP header 'Accept' should have a specific type `application/vnd+amazon.uitoolkit+json`. | Skill list | SKILL | {<br>  "title": "*skill title*",<br>  "productDetails": { "releaseDate": "1456958015" },<br>  "developerInfo": { "name": "*developer name*" },<br>  "entitlementInfo": { "accountLinked": true }<br>} |
| Compatible Device | https://pitangui.amazon.com/api/**phoenix** | Detected compatible devices | COMPATIBLE_DEVICE TIMELINE | "uuid:*uuid*": {<br>  "modelName": "*model name*",<br>  "friendlyName": "white lamp 1",<br>  "friendlyNameModifiedAt": 1481558860291,<br>  "applianceNetworkState": {<br>    "createdAt": 1481558860291<br>  }<br>} |
| User Activity | https://pitangui.amazon.com/api/**todos**?type=TASK&size={} | To-do list | TIMELINE | {<br>  "createdDate": 1480350314486,<br>  "lastupdatedDate": 1480350314486,<br>  "customerId": "\*\*E99\*\*\*0L\*\*\*Z",<br>  "originalAudioId": "*URL of the voice file on the cloud*",<br>  "text": "do the laundry"<br>} |
| User Activity | https://pitangui.amazon.com/api/**todos**?type=SHOPPING_ITEM&size={} | Shopping list | TIMELINE | |
| User Activity | https://pitangui.amazon.com/api/**notifications** | Timer and alarm list | TIMELINE | {<br>  "alarmTime": 1480363200000,<br>  "createdDate": 1480355912857,<br>  "deviceSerialNumber": "\*\*90\*\*\*964\*\*\*\*\*U",<br>  "status": "ON", "type": "Alarm"<br>} |
| User Activity | https://pitangui.amazon.com/api/**cards**?beforeCreationTime={} | Card list (conversations between users and Alexa) | TIMELINE | {<br>  "cardType": "TextCard",<br>  "creationTimestamp": 1484539461678,<br>  "playbackAudioAction": {<br>    "mainText": "*text what Alexa heard*",<br>    "url": "*URL of the voice file on the cloud*"<br>  },<br>  "sourceDevice": { "serialNumber": "\*\*90\*\*\*964\*\*\*\*\*U" },<br>  "title": "Do you know the muffin man?"<br>} |
| User Activity | https://pitangui.amazon.com/api/**activities**?startTime={}&size={}&offset=-1 | History on voice interactions with Alexa | TIMELINE | {<br>  "activityStatus": "SUCCESS",<br>  "creationTimestamp": 1484542204396,<br>  "registeredCustomerId": "\*\*E99\*\*\*0L\*\*\*Z",<br>  "sourceDeviceIds": [{<br>    "serialNumber": "\*\*90\*\*\*964\*\*\*\*\*U"<br>  }],<br>  "utteranceId": "*URL of the voice file on the cloud*"<br>} |
| User Activity | https://pitangui.amazon.com/api/**media**/historical-queue?deviceSerialNumber={}&deviceType={}&size={}&offset=-1 | Music Playing list | TIMELINE | {<br>  "providerId": "*provider*",<br>  "startTime": 1484542090384,<br>  "title": "80s 90s & Today - \*\*\*\*\*"<br>} |
| ETC | https://pitangui.amazon.com/api/**utterance**/audio/data?id={*originalAudioId* or *utteranceId*} | Accessing to audio data (actual user's voice) | - | - |

Appendix B. Examples of a normalized database created by Alexa modules of *CIFT*

| Table name | Column name | Example data (Personal info. is masked by asterisks and italic text) | Description |
|---|---|---|---|
| ACQUIRED_FILE (File list stored at the evidence library) | id | 1 | Unique ID |
| | operation | CLOUD | Operation type |
| | src_path | https://pitangui.amazon.com/api/bootstrap | URL or file path |
| | desc | A Bootstrap Account | Description |
| | saved_path | …*(omitted)*.../a57665bd28e72163adb0245b51ac6fb02e45ed44.json | File path saved in the evidence library |
| | sha1 | fffe474050329d0bb0691074d401b7a324a1b9a2 | SHA-1 hash value of raw data |
| | saved_timestamp | 2017-01-29 20:56:27 | Saved timestamp |
| | timezone | *time zone* | Local time zone |
| ACCOUNT | customer_email | \*\*n\*h\*\*@\*mail.com | User account ID |
| | customer_name | *First Last* | User name |
| | customer_id | \*\*E99\*\*\*\*0L\*\*\*Z | Customer ID |
| | source_id | 1 | Foreign key (ACQUIRED_FILE) |
| ALEXA_DEVICE (Alexa-enabled devices) | device_account_name | TEEST's 2nd Echo Dot | Device account name |
| | device_account_id | \*\*C10\*\*\*ND\*\*\*W | Device account ID |
| | customer_id | \*\*E99\*\*\*\*0L\*\*\*Z | Customer ID |
| | device_serial_number | \*\*\*0L9\*\*\*473\*\*\*P | Device serial number |
| | device_type | A3S5BH2HU6VAYF | Echo's device type |
| | sw_version | 564197320 | Software version |
| | mac_address | \*\*63\*\*6C\*\*D5 | Mac address |
| | address | *address* | Home Address |
| | postal_code | *zipcode* | Postal code |
| | locale | *locale* | Locale |
| | timezone | *time zone* | Time zone |
| | source_id | 9 | Foreign key (ACQUIRED_FILE) |
| SETTING_WIFI | ssid | *SSID* | SSID |
| | security_method | WPA_PSK | Security method |
| | pre_shared_key | *password* | Wi-Fi password (plain text) |
| | source_id | 3 | Foreign key (ACQUIRED_FILE) |
| SETTING_MISC (Misc settings) | name | calendar_account | Name of the value |
| | value | { "emailId": "\*\*\*al\*\*un\*@\*mail.com", "calendarList": [{ "calendarId": "*calendar id*", "calendarName": "*name*" }] } | Account name / Calendar ID / Calendar name |
| | source_id | 5 | Foreign key (ACQUIRED_FILE) |
| SKILL | title | NASA Mars | Skill title |
| | developer_name | Jet Propulsion Laboratory | Developer name |
| | account_linked | False | True if a user account is linked |
| | release_date | 2016-11-29 20:05:25.000 | Release date of this skill |
| | source_id | 61 | Foreign key (ACQUIRED_FILE) |
| COMPATIBLE_DEVICE (Compatible devices around Alexa devices) | name | \*\*\*\* Switch | Device name |
| | manufacture | *manufacture* | Manufacture name |
| | model | Socket | Model name |
| | created | 2017-01-26 23:18:14.584 | First detected time |
| | name_modified | 2017-01-26 23:18:14.584 | Friendly name changed time |
| | desc | \*\*\*\* Switch | Description |
| | type | urn:\*\*\*\*\*\*:device:controllee:1 | Device type |
| | reachable | True | Reachable or not |
| | firmware_version | \*\*\*\*\_\*\*\_\*.00.10885.\*\*\* | Firmware version |
| | appliance_id | uuid:Socket-1_0-\*\*\*\*26\*\*\*008AD | Appliance ID |
| | alexa_device_serial_number | \*\*90\*\*\*964\*\*\*\*\*U | Device serial number |
| | alexa_device_type | A3S5BH2HU6VAYF | Device type (it means Echo) |
| | source_id | 11 | Foreign key (ACQUIRED_FILE) |
| TIMELINE (An integrated table for information with timestamps) | date | 2017-01-17 | Date |
| | time | 08:10:44.175 | Time |
| | timezone | *time zone* | Time zone |
| | MACB | …B | MACB time format |
| | source | CLOUD | Operation type |
| | sourcetype | Activity History | Detailed source type ('Home' menu) |
| | type | Created | Type of timestamp |
| | user | \*\*E99\*\*\*\*0L\*\*\*Z | Customer ID |
| | host | \*\*90\*\*\*964\*\*\*\*\*U | Serial number |
| | short | History | Google calendar card |
| | desc | add driver license to to do list | Google calendar |
| | filename | …*(omitted)*.../7d40452a7ba57fd4f91a248f561cbf77dcb70882.json | File path saved in the evidence library |
| | notes | SUCCESS | Text what Alexa heard |
| | format | JSON | JSON format |
| | extra | User's voice: "https://pitangui.amazon.com/api/utterance/audio/data?id=A3S5BH2HU6VAYF:1.0/2017/01/17/13/\*\*90\*\*\*964\*\*\*\*\*U/10:38::TNIH_2V.f78ee15e-ef2e-45a8-890f-690ef5e543abZXV" | URL of the voice file on the cloud |